\documentclass[lettersize,journal]{IEEEtran}
\usepackage{amsmath,amsfonts}
\usepackage{algorithmic}
\usepackage{algorithm}
\usepackage{array}
\usepackage[caption=false,font=normalsize,labelfont=sf,textfont=sf]{subfig}
\usepackage{textcomp}
\usepackage{stfloats}
\usepackage{url}
\usepackage{verbatim}
\usepackage{graphicx}
\usepackage{cite}
\usepackage[hidelinks]{hyperref}
\usepackage{amsthm}
\theoremstyle{definition}
\newtheorem{definition}{Definition}
\usepackage{amssymb}

\usepackage{xcolor}
\usepackage{listings}

\definecolor{codeblue}{RGB}{0,0,170}
\definecolor{codered}{RGB}{170,0,0}
\definecolor{codegray}{RGB}{100,100,100}

\lstdefinelanguage{Solidity}{%
    keywords={
        function, contract, mapping, address,
        uint, uint256, private, public,
        internal, external, returns, return,
        emit, event, if, else, for, while,
        memory, storage, calldata, payable
    },
    sensitive=true,
    morecomment=[l]{//}
}

\lstdefinestyle{soliditycompact}{%
    language=Solidity,
    basicstyle=\ttfamily\small\color{black},
    keywordstyle=\color{codeblue}\bfseries,
    commentstyle=\color{codegray}\itshape,
    emph={userDebts},
    emphstyle=\color{codered}\bfseries,
    columns=fullflexible,
    keepspaces=true,
    showstringspaces=false,
    breaklines=true,
    breakatwhitespace=true,
    tabsize=2,
    frame=single,
    framerule=0.4pt,
    framextopmargin=0.3em,
    framexbottommargin=0.3em,
    aboveskip=6pt,
    belowskip=6pt,
    captionpos=b,
    numbers=none
}

\usepackage{booktabs}
\usepackage{multirow}
\usepackage{array}
\usepackage{tabularx}
\usepackage{ragged2e}
\usepackage{makecell}
\usepackage{orcidlink}

\begin{document}

\title{Tracing Money Laundering Across DeFi, Bridges, and Mixers on Ethereum}

\author{Huan~Xing,
Gaohe~Jin, Bang~Fu, Yue~Li, Baoning~Niu, and Xing~Fan
\thanks{Huan Xing, Gaohe~Jin, Bang~Fu, Yue Li, Baoning Niu, and Xing Fan are with the College of Computer Science and Technology (College
of Data Science), Taiyuan University of Technology, Jinzhong, Shanxi 030600, China (e-mail: xinghuan@tyut.edu.cn; 2024520658@link.tyut.edu.cn; 2024319014@link.tyut.edu.cn; liyue@tyut.edu.cn; niubaoning@tyut.edu.cn; fanxing@tyut.edu.cn).}%
\thanks{Corresponding author: Yue Li.}
}

\markboth{Journal of \LaTeX\ Class Files,~Vol.~14, No.~8, August~2021}%
{Shell \MakeLowercase{\textit{et al.}}: A Sample Article Using IEEEtran.cls for IEEE Journals}


\maketitle

\begin{abstract}
Illicit actors exploit Ethereum's deep liquidity and diverse on-chain services to launder criminal proceeds by combining decentralized finance (DeFi) protocols, cross-chain bridges, and mixers in multi-stage processes. Existing anti-money laundering tracing methods fail to disentangle the underlying asset-flow paths in complex transactions, potentially misdirecting subsequent tracing; cannot identify cross-chain intent without pre-collected bridge information; and treat mixers as generic exit services, overlooking post-mixer laundering flows. We present FPTracer, a framework for tracing illicit assets across Ethereum services. First, we propose a method based on the Context-Aware Transfer Tree (CATT) to recover intra-transaction asset-flow paths, which guide accurate tracing and service analysis. Second, we extract discriminative features from asset-flow and execution-evidence perspectives and train a classifier to identify transactions with cross-chain intent. Third, we leverage evidence from the upstream laundering network to analyze potential post-mixer laundering activities. Evaluation on 20 real-world laundering cases demonstrates FPTracer's effectiveness. It achieves the highest recall for both known laundering addresses (98.6\%) and destinations (94.2\%). It identifies all known uses of bridges in these cases and uncovers 30 previously unreported ones. Among 18 cases with identified Tornado Cash usage, FPTracer can identify potential withdrawal-side activities, which are corroborated by public reports where available.
\end{abstract}


\section{Introduction}
\begin{figure}[!t]
  \centering
\includegraphics[width=\columnwidth]{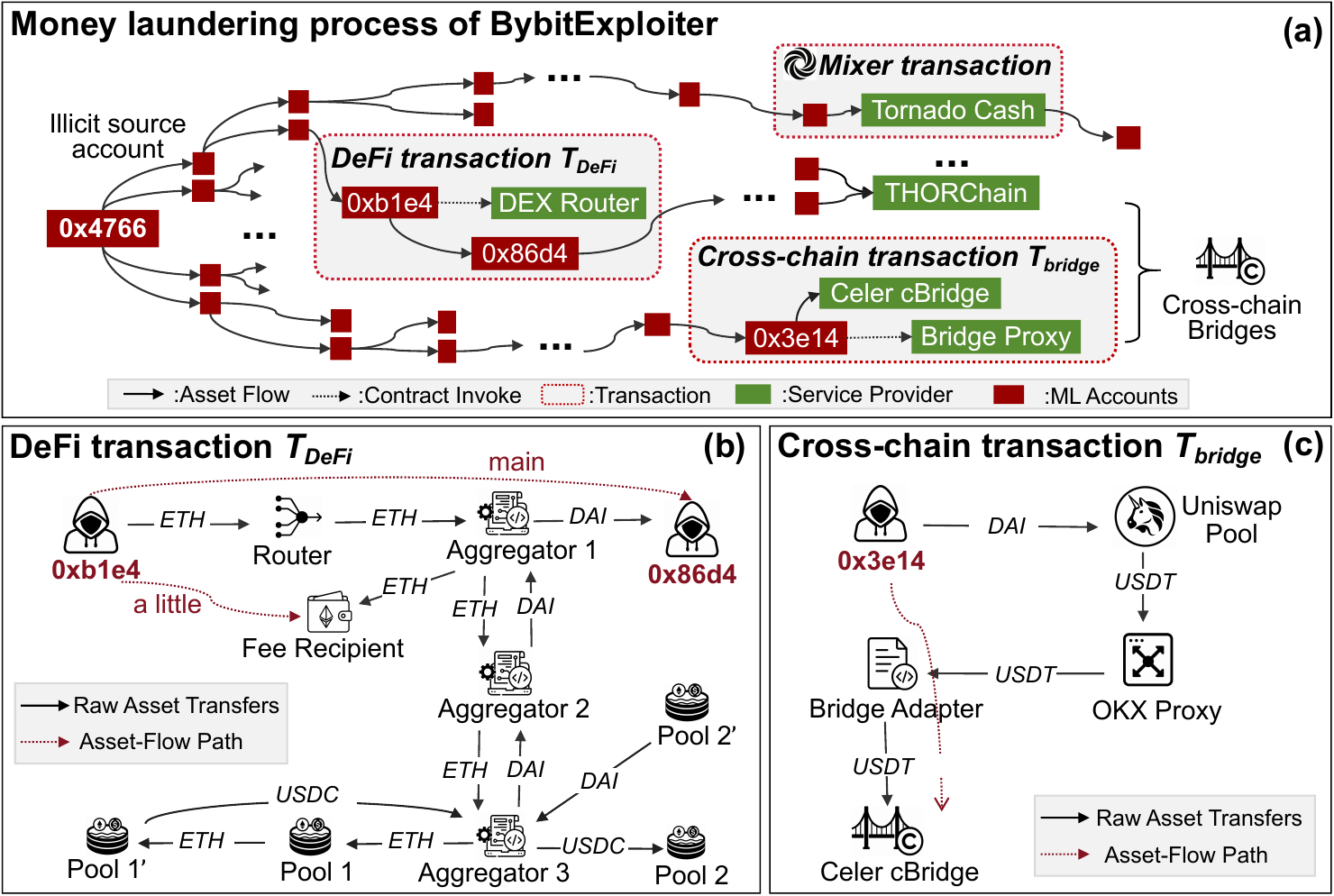}
\caption{
Illustration of the BybitExploiter laundering process.
(a) Simplified laundering network.
(b) A DeFi transaction \(T_{\mathrm{DeFi}}\) involved in money laundering, containing multiple raw asset transfers that obscure the underlying asset-flow paths.
(c) A cross-chain transaction \(T_{\mathrm{deposit}}\) involved in money laundering, in which illicit assets pass through intermediary contracts before finally reaching the cross-chain bridge.
} 
\label{fig:example}
\end{figure}

\IEEEPARstart{T}{he} pseudonymity and permissionless nature of blockchains have made cryptocurrencies attractive to illicit actors seeking to launder proceeds from a wide range of criminal activities~\cite{Manning2025CryptoML,FATF2025VA}.
Money laundering (ML) aims to conceal the criminal origin
of these proceeds and make them appear legitimate.
Unlike traditional UTXO-based blockchains such as Bitcoin, Ethereum supports programmable smart contracts, enabling a rich ecosystem of on-chain services~\cite{EthereumApps}.
These services can be exploited to facilitate sophisticated money laundering activities~\cite{FATF2026DeFi}.
Specifically, decentralized finance (DeFi) protocols provide permissionless financial services for exchanging, lending, and managing cryptoassets. 
A single transaction in DeFi may involve complex asset-flow logic, such as repeated asset conversions and routing through multiple service-related intermediary addresses, which are not explicitly presented in the resulting raw asset transfers.
Cross-chain bridges allow assets to move from Ethereum to other blockchains, causing asset flows to span multiple independent ledgers. Mixers introduce an additional layer of obfuscation by using cryptographic techniques to break the observable link between incoming and outgoing asset flows.
Illicit actors often combine several of these services across multiple stages to progressively obscure the flow of illicit assets before ultimately converting them into fiat currency through off-ramp services~\cite{Elliptic2021DeCrime}.
By 2025, more than \$21.8 billion in cryptoassets had been laundered using cross-chain methods~\cite{Elliptic2025CrossChainCrime}, while approximately \$655 million in cryptoassets associated with illicit activity flowed into mixers in 2025 alone~\cite{GlobalLedger2026Laundering}.
On February 21, 2025, cryptocurrency exchange Bybit suffered the largest reported theft to date, resulting in losses of approximately \$1.46 billion~\cite{Elliptic2025Typologies}. 
Fig.~\ref{fig:example}(a) presents a simplified illustration of the subsequent laundering process.
After the theft, the hackers consolidated the stolen asset portfolio (e.g., ETH, stETH, cmETH, mETH) into a self-controlled account \texttt{0x4766\dots} and initiated the laundering by dispersing stolen assets across numerous intermediary accounts through multi-hop transfers.
During this process, they used decentralized exchanges (DEXs), a DeFi service, to perform asset conversions, such as swapping stETH for the more liquid ETH, thereby facilitating subsequent laundering and reducing the risk of asset freezing.
The stolen assets were subsequently laundered through cross-chain transfers from Ethereum to other blockchains (e.g., Bitcoin) and through mixers such as Tornado Cash, further obscuring their flows.

Anti-money laundering (AML) constitutes a critical last line of defense against various criminal activities.
Numerous AML methods are proposed to track the flow of illicit assets, including their movement and destinations, providing actionable evidence to support subsequent investigations, asset freezing, and recovery. 
Existing tracing approaches~\cite{Wu2024,wu2023tracer,wu2026ettracker,wu2026tracingshadowsautomatictracking} primarily rely on address-level relationships derived from raw asset transfers or inferred transaction semantics.
They do not explicitly disentangle underlying asset-flow paths within complex transactions, limiting their ability to accurately trace downstream recipients of illicit assets.
This limitation may introduce noise into laundering traces and miss actual laundering flows.
Meanwhile, they either fail to identify whether illicit assets enter a cross-chain transfer process or only identify those entering bridges covered by pre-collected information.
As a result, the cross-chain intent may go unrecognized, leaving no concrete evidence for regulators or security practitioners to conduct subsequent analysis on other blockchains.
Furthermore, they take mixers as generic exit services, leaving potential post-mixer laundering flows on Ethereum unanalyzed.
Overall, these limitations expose three key challenges in laundering tracing on Ethereum.

\textbf{C1: Implicit asset flows in complex Ethereum transactions.}
Complex Ethereum transactions involving on-chain services may contain nontrivial asset-processing logic. 
In the Bybit case, the hackers initiated a DeFi transaction%
\footnote{\href{https://etherscan.io/tx/0xc2dbe6cedf21843f3b94fe6a72ecc8a324d0e39cfe91ba9215723681e484386a}{Laundering transaction: \texttt{0xc2db...386a}}}
to swap assets.
As shown in Fig.~\ref{fig:example}(b), such a transaction generates multiple raw token transfers, while the tracked assets are actually routed through several intermediate contracts (e.g., router or aggregation contracts, liquidity pools) and converted into another asset before reaching their downstream recipient.
These raw transfers obscure the underlying asset-flow paths, making the actual downstream recipient difficult to determine.
However, existing approaches fail to accurately recover such paths within individual transactions, which may result in misdirected or incomplete laundering tracing.

\textbf{C2: Elusive Cross-Chain Intent.}
Once tracked illicit assets are directed into cross-chain bridges to initiate transfers to other blockchains, continued tracing on Ethereum becomes useless and misleading.
Identifying such cross-chain intent is therefore crucial for tracing illicit asset flows.
However, cross-chain bridges adopt diverse designs and exhibit no unified on-chain interaction pattern: they share no common interface or standardized event signature, and their transactions may resemble ordinary DeFi interactions.
In practice, some cross-chain interactions are even routed through proxy contracts rather than directly invoking the underlying bridge logic, where asset conversion and bridge entry may occur within the same transaction, as illustrated by the transaction%
\footnote{\href{https://etherscan.io/tx/0x09d0b52446e9d23753cedaa7dc1ef49532767e22e579cbc3b8998aefe082434d}{Laundering transaction: \texttt{0x09d0...434d}}} in Fig.~\ref{fig:example}(c).
Consequently, determining whether transactions involving tracked illicit assets exhibit cross-chain intent remains challenging.

\textbf{C3: Protocol-enforced tracing discontinuity at mixers.}
Mixers use cryptographic mechanisms to break the direct on-chain linkage between assets entering and leaving the service.
As a result, once illicit assets enter a mixer, their subsequent movement cannot be directly traced, disrupting the continuity of laundering process.

To address these challenges, we develop FPTracer (\textbf{F}low-\textbf{P}ath \textbf{Tracer}), an AML tracing framework on Ethereum. 
FPTracer uses the underlying asset-flow paths implicit in raw asset transfers within transactions to guide iterative tracing of illicit assets and construct the laundering network. It further identifies the involvement of cross-chain bridges and mixers in laundering activities, avoiding premature termination of tracing at these services and preserving evidence for subsequent analysis.
Our main contributions are summarized as follows:

\begin{itemize}
\item To address \textbf{C1}, we introduce a transaction-level analysis method based on the Context-Aware Transfer Tree (CATT), a data structure that integrates raw asset transfers with transaction execution structure. 
This method recovers underlying asset-flow paths within each transaction, without requiring prior knowledge of contract-specific operations.
The recovered paths enable more accurate tracing of illicit assets  and facilitate the analysis of involved services.

\item To address \textbf{C2}, we capture general cross-chain behavioral patterns from asset flows and transaction execution, derive discriminative features from these patterns, and use a learning-based classifier to identify transactions that exhibit cross-chain intent. This design avoids reliance on manually curated bridge-specific information.

\item To address \textbf{C3}, we identify when tracked assets enter a mixer and leverage evidence accumulated in the preceding laundering network to flag potential post-mixer laundering activities.

\item We conduct a comprehensive evaluation of FPTracer on 20 real-world laundering cases from 2019 to 2025, involving approximately \$5.47 billion in reported losses. 
On the 10 cases with labeled laundering traces, FPTracer achieves the highest recall among the compared methods, identifying 98.6\% of labeled laundering addresses and 94.2\% of labeled laundering destinations.
Across all 20 cases, FPTracer identifies all publicly reported bridge uses and discovers 30 additional uses not documented in public reports.
FPTracer further finds Tornado Cash usage in 18 cases and identifies potential withdrawal-side laundering activities in 7 cases. For the KucoinHacker and LiquidHacker cases, we find public reports corroborating 222 of 442 and 47 of 82 candidate withdrawal transactions, respectively.

\end{itemize}

\section{Background}
\label{sec2}
\subsection{Transaction Execution Model on Ethereum.}
Ethereum has two types of accounts: externally owned accounts (EOAs), which are controlled by private keys, and contract accounts (CAs), which execute predefined logic when invoked.
Each account is associated with a unique identifier, referred to as an address.
A transaction is a signed instruction initiated by an EOA, denoted by \texttt{tx.from}, to a recipient, denoted by \texttt{tx.to}. The transaction initiates a message call from \texttt{tx.from} to \texttt{tx.to}, which constitutes the top-level call.
If the recipient is a CA, the call executes its code and may further triggers other contracts through nested internal message calls within the same transaction. These calls form a hierarchical execution trace according to their nesting relationships and execution order.
Asset transfers may occur during transaction execution, involving either the native asset ETH or tokens managed by contracts. 
ETH can be transferred through message calls carrying nonzero value. 
Token transfers are executed by contracts, following standards such as ERC-20 and ERC-721, and are typically reported through standardized \texttt{Transfer} events recorded in transaction logs.
Beyond asset transfers, contract execution may emit other event logs indicating specific operations and modify persistent contract storage. Such storage modifications produce state changes that reflect updates to the contract's on-chain state.

\subsection{On-chain Service Providers.}
The Ethereum ecosystem provides a wide range of financial and technical on-chain services.
We refer to the entities or protocols providing these services as service providers (SPs). Unlike ordinary recipients, SPs may receive assets as part of specific service interactions and therefore require different treatment in laundering tracing. We consider four major types of SPs involved in laundering activities.

\textbf{DeFi protocols.}
Decentralized finance (DeFi) provides financial services through smart contracts without centralized intermediaries.
Typical DeFi services include decentralized exchanges (DEXs), which enable users to trade different tokens; lending protocols, which allow users to deposit collateral to borrow other assets; and staking or yield protocols, which allow users to supply assets to earn rewards. 

\textbf{Cross-chain bridges.}
\label{bridge_background}
Cross-chain bridges enable asset transfers between otherwise isolated blockchains.
Although their implementations vary, a cross-chain transfer generally involves coordinated operations on a source chain and a destination chain, together with a mechanism for cross-chain information propagation and verification~\cite{ren2023interoperability}.
On the source chain, users initiate a deposit transaction whose execution transfers assets to a bridge contract, where the assets may be locked or burned.
The execution may produce observable on-chain information, such as event logs, for subsequent cross-chain processing.
Off-chain components propagate this information to the destination chain and, depending on the bridge design, may also participate in its validation. On the destination chain, the bridge contract verifies the received information and then releases or mints the corresponding assets to the specified recipient.

\begin{figure}[!t]
\includegraphics[width=\columnwidth]{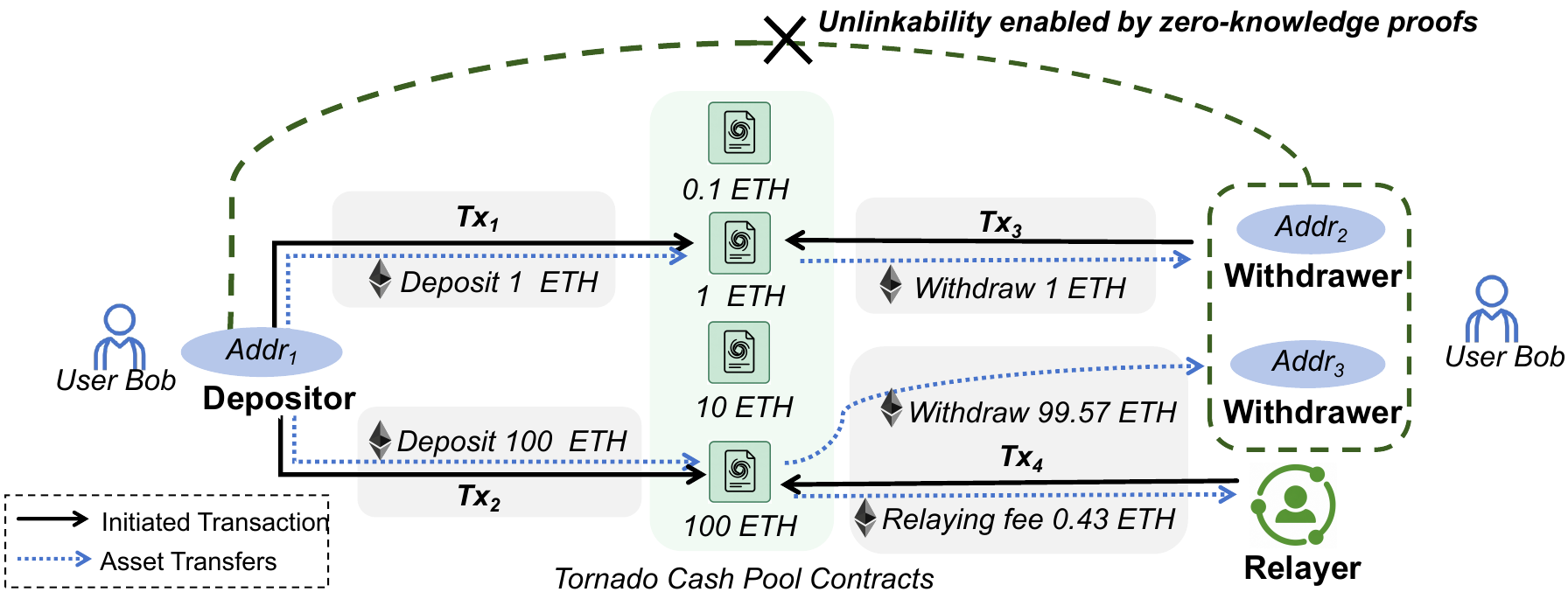}
\caption{The process of user interaction with Tornado Cash.} 
\label{fig:tornado}
\end{figure}

\textbf{Mixers.}
\label{tornado_background}
Mixers are privacy-enhancing services that allow users to mix their assets with those of other users in order to break the linkability between addresses. Tornado Cash (TC), a widely used non-custodial mixer on Ethereum, achieves this through zero-knowledge proofs.
It operates four fixed-denomination ETH pool contracts for 0.1, 1, 10, and 100 ETH.
In a typical workflow, a user deposits a fixed amount of ETH into the corresponding pool from a \emph{depositor} address and keeps a private note. The note is later used to generate a zero-knowledge proof for withdrawal. After verifying the proof, the pool contract transfers the corresponding ETH to a user-specified recipient address, referred to as the \emph{withdrawer}. 
Tornado Cash also supports a relayer service.
The withdrawal transaction can be submitted by a \emph{relayer}, which pays the gas fee and receives a relaying fee, while the withdrawn assets are still sent to the withdrawer.
For example, in Fig.~\ref{fig:tornado}, a user Bob aims to conceal the trail of 101 ETH under his control.
He first uses \(Addr_1\) to initiate two deposit transactions, \(Tx_1\) and \(Tx_2\), depositing 1 ETH and 100 ETH into the corresponding pools, respectively.
He later uses a fresh address, \(Addr_2\), to submit \(Tx_3\) and withdraw 1 ETH.
For the remaining 100 ETH, Bob uses a relayer to submit the withdrawal transaction \(Tx_4\) on his behalf. The relayer receives a fee of 0.43 ETH, while the remaining 99.57 ETH is transferred to \(Addr_3\), another address controlled by Bob. 
Although these deposit and withdrawal transactions are observable on-chain, Tornado Cash break the direct on-chain link between the depositor address (eg., \(Addr_1\)) and the withdrawal addresses (eg., \(Addr_2\) and \(Addr_3\)).

\textbf{Centralized exchanges (CEXs).}
CEXs are custodial platforms that allow users to trade assets and convert them into fiat currency.
In laundering workflows, they often serve as exit services for trading or cash-out.
Once assets enter a CEX, further investigation generally shifts off-chain and relies on regulatory cooperation and Know Your Customer (KYC) information.


\section{Methodology of FPTracer}
\label{sec3}

\subsection{Overview}
\begin{figure*}[!t]
    \centering
    \includegraphics[width=\textwidth]{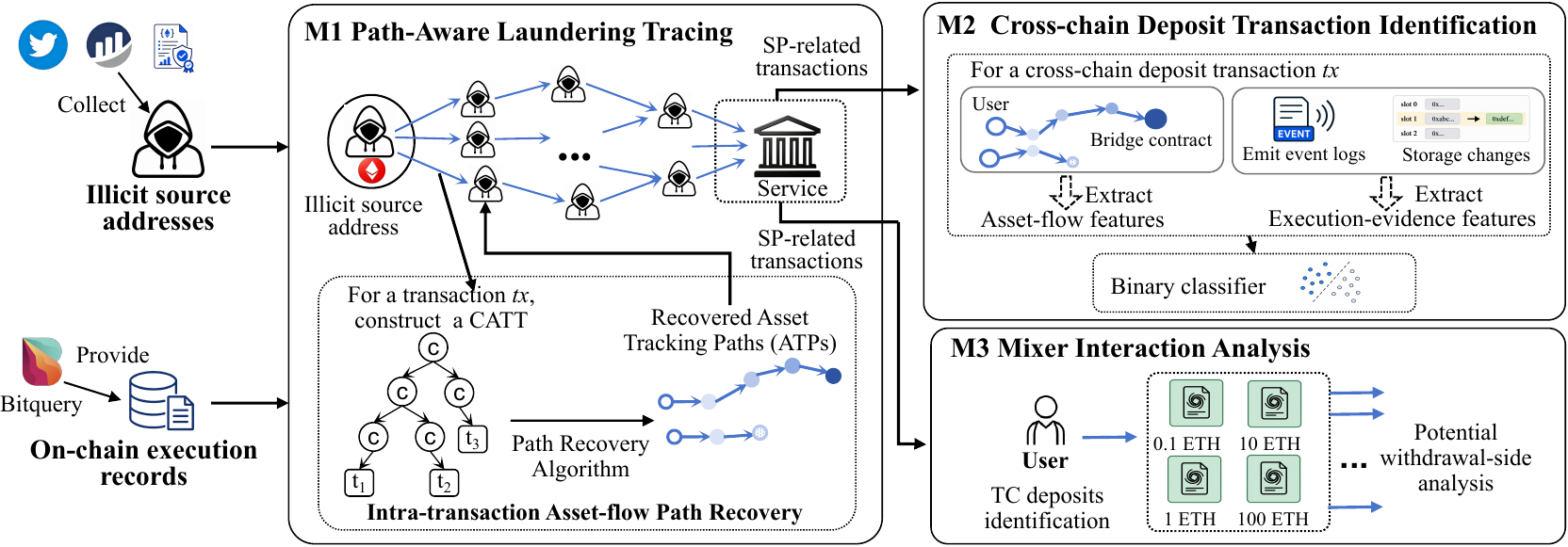}
    \caption{Overview of FPTracer.}
    \label{fig1}
\end{figure*}

This paper presents FPTracer, an AML tracing framework for Ethereum, as shown in Fig.~\ref{fig1}.
Starting from known illicit source addresses, FPTracer traces the movement of illicit assets by analyzing the underlying asset-flow paths within transactions, incrementally constructs the laundering network, and identifies the use of cross-chain bridges and mixers.
First, FPTracer performs transaction-level analysis to decipher how tracked assets propagate and transform among relevant addresses and accurately identify their final recipients within a transaction.
Specifically, \textbf{M1} constructs a Context-Aware Transfer Tree (CATT) for each transaction to organize raw asset transfers according to their positions and relationships in the execution trace. It then recovers the underlying asset-flow paths, represented as Asset Tracking Paths (ATPs), based on structural constraints and value continuity.
FPTracer then follows the destinations of ATPs associated with the tracked assets to iteratively expand the laundering network. When the tracked assets reach a service-provider boundary, regular network expansion stops and, where applicable, the corresponding service-specific analysis is invoked.
In \textbf{M2}, FPTracer further determines whether the corresponding transaction represents a cross-chain bridge deposit. It extracts effective features from asset flows and execution evidence, such as event logs and storage changes, and applies a learned classifier for identification.
In \textbf{M3}, FPTracer first identifies Tornado Cash deposit transactions from the destinations of relevant ATPs. 
It then leverages Tornado Cash's implementation details and evidence from the preceding laundering network to design rules for flagging potential withdrawal-side laundering transactions.

\subsection{Path-Aware Laundering Tracing}

\subsubsection{Intra-transaction Asset-flow Path Recovery}
\label{subsec4.1}
A single Ethereum transaction may generate multiple raw asset transfers. These interleaved transfers span multiple addresses and asset types, obscuring the underlying value flow and making illicit asset tracing difficult.
To address this problem, we propose a novel data structure, the CATT, and a path recovery algorithm to find the underlying asset-flow paths.

\paragraph{CATT Construction} CATT associates each raw asset transfer with the call context in which it occurs, thereby providing structural constraints for subsequent transfer association and preventing unrelated transfers from being incorrectly linked. 
For each transaction, we collect the message calls with nesting relationships and execution order, together with asset transfers, from the execution trace and transaction logs to construct CATT.

\begin{definition}[Context-Aware Transfer Tree]
\label{CATT}
For a transaction $tx$, a CATT $\mathcal{T} = (V_C, V_A, E, \prec)$ is an ordered tree. \(V_C\) is the set of call nodes, including the message calls triggered during the execution. 
\(V_A\) is the set of asset transfer nodes, representing ETH and token transfers.
\(E\) is the edge set, containing two types:    
a call edge \((u, v)\), where \(u, v \in V_C\), indicates that call \(u\) invokes call \(v\);
a transfer edge \((u, t)\), where \(u \in V_C\) and \(t \in V_A\), indicates that transfer \(t\) is triggered during the execution of the call \(u\).
The relation $\prec$ specifies the execution order of sibling nodes. For two nodes $u$ and $v$ with the same parent, $u \prec v$ indicates that \(u\) is executed before \(v\).
\end{definition}

\begin{figure*}[!t]
  \centering
\includegraphics[width=\textwidth]{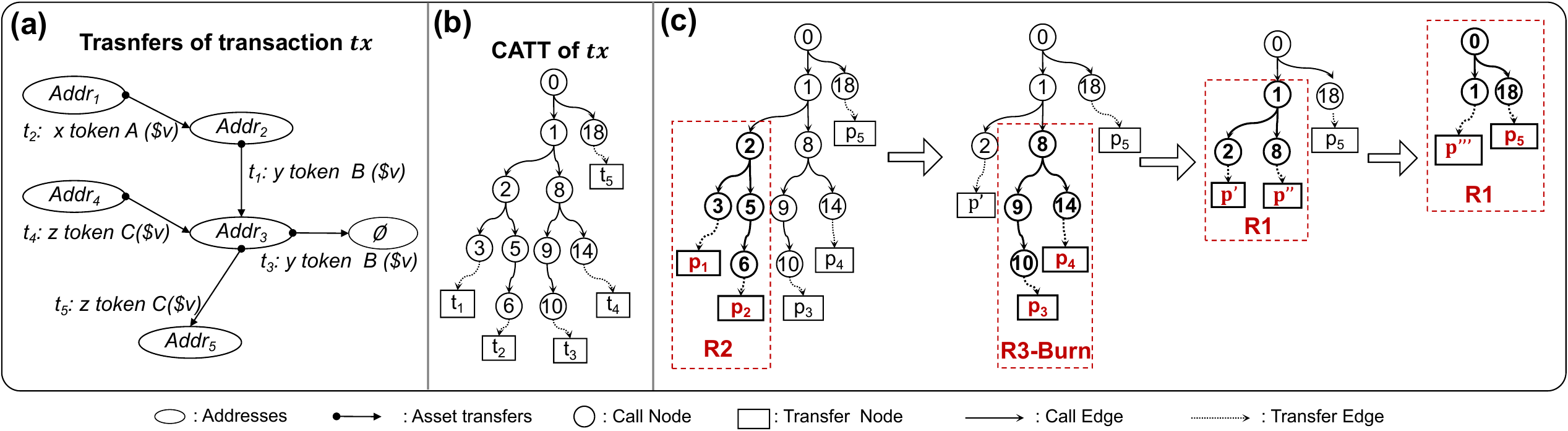}
\caption{
Illustration of CATT and asset-flow path recovery.
    (a) An transaction example $tx$ containing five raw asset transfers ($t_1$--$t_5$). For example, $t_2$ transfers $x$ units of token A from $Addr_1$ to $Addr_2$, with a USD value of $v$.
    (b) The CATT constructed from the message calls and asset transfers of $tx$.
    (b) Path recovery process based on the CATT. Each transfer node is first initialized as an atomic ATP. These atomic ATPs are progressively associated during a postorder traversal of the CATT by applying three path-association rules.
} 
\label{fig:catt}
\end{figure*}

We illustrates the construction of CATT in three steps.
First, we instantiate all message calls as call nodes. Their nesting relationships are then used to construct call edges by linking each call to its parent, while the sibling execution order is used to establish \(\prec\).
Second, we instantiate asset transfers, including ETH and token transfers, as transfer nodes and connect each transfer node to the call node in which it occurs through a transfer edge.
Finally, to eliminate noise from auxiliary operations (e.g., \texttt{approve} and \texttt{isOwner}), we prune call nodes that neither trigger asset transfers nor have descendants that trigger asset transfers.
An example CATT constructed for transaction \(tx\) is shown in Fig.~\ref{fig:catt}(b). The number in each call and transfer node denotes its execution order in the trace. Because calls that neither directly nor indirectly trigger asset transfers are pruned, the remaining call indices may be non-consecutive. 

\paragraph{Path Recovery Algorithm}
Given a CATT, the next step is to recover the underlying asset-flow paths by associating transfers that belong to the same value flow.
To support this process, we define the ATP to model the recovered or partially recovered asset-flow path within a transaction.
An ATP may contain multiple sources and destinations, allowing it to represent split and merged asset flows rather than only linear transfer chains.

\begin{definition}[Asset Tracking Path]
An ATP is a tuple $p=\langle \mathcal{E},\mathcal{S},\mathcal{D} \rangle$, where $\mathcal{E}$ records the asset transfers forming \(p\), and the set $\mathcal{S}$ and $\mathcal{D}$ record the source and destination entries of the recovered asset flow, respectively. 
Each entry $\langle a^s,\tau^s,q^s,v^s\rangle \in \mathcal{S}$ represents an amount $q^s$ of token \(\tau^s\), valued at $v^s$ USD, that originates from address $a^s$. 
Each entry $\langle a^d,\tau^d,q^d,v^d\rangle \in \mathcal{D}$ represents an amount \(q^d\) of token \(\tau^d\) , valued at \(v^d\) USD, that reaches address \(a^d\).  
\end{definition}

\begin{algorithm}[!t]
\caption{Path Recovery Algorithm}
\label{alg:bottomup-recovery}
\footnotesize
\begin{algorithmic}[1]
\REQUIRE A CATT $\mathcal{T}=(V_C,V_A,E,\prec)$
\ENSURE A recovered ATP set $\mathcal{P}$

\FOR{each transfer node $t \in V_A$}
    \STATE $p \gets \textsc{InitAtomicPath}(t)$
    \label{alg:line:init-path}
    \STATE $\textsc{ReplaceTransferWithPath}(t,p)$
    \label{alg:line:replace-transfer}
\ENDFOR

\FOR{each call node $c \in \textsc{PostorderTraverse}(\mathcal{T})$}
    \STATE $\mathcal{P}_c \gets \textsc{CollectPathsInOrder}(c,\prec)$
    \label{alg:line:collect-paths}

    \IF{$|\mathcal{P}_c| \geq 2$}
        \STATE $\mathcal{P}_c \gets \textsc{AssociatePaths}(\mathcal{P}_c,c)$
        \label{alg:line:associate-paths}
    \ENDIF

    \STATE $\textsc{ReplaceDescendantsWithPaths}(c,\mathcal{P}_c)$
    \label{alg:line:replace-descendants}
\ENDFOR

\STATE $\mathcal{P} \gets \textsc{GetRootPathSet}(\mathcal{T})$
\label{alg:line:get-root}

\STATE \textbf{return} $\mathcal{P}$
\label{alg:line:return}
\end{algorithmic}
\end{algorithm}

We recover the ATP set of a transaction from the CATT using a bottom-up procedure shown in Algorithm~\ref{alg:bottomup-recovery}.
First, each transfer node is initialized as an atomic ATP and replaces the original transfer node in the CATT (Lines~\ref{alg:line:init-path}--\ref{alg:line:replace-transfer}).
The algorithm then traverses all call nodes in postorder. 
For each call node $c$, it collects the ATPs currently retained in its subtree according to the execution order $\prec$, yielding \(\mathcal{P}_c\) (Line~\ref{alg:line:collect-paths}).
If \(\mathcal{P}_c\) contains at least two ATPs, the algorithm invokes \textsc{AssociatePaths} to associate and merge ATPs according to the defined rules (Line~\ref{alg:line:associate-paths}).
The current descendants of $c$ are then replaced by the resulting ATP sequence (Line~\ref{alg:line:replace-descendants}).
After the root call node is processed, the ATPs retained under the root form the recovered ATP set $\mathcal{P}$ (Lines~\ref{alg:line:get-root}--\ref{alg:line:return}).
By processing deeper call contexts first, the postorder traversal prioritizes associations within local execution contexts and avoids premature associations across unrelated contexts.

Prior studies have shown that, despite their diverse business logic, DeFi applications exhibit recurring financial behaviors from an asset-flow perspective~\cite{LLMDefiPrice,Li2023}.
Motivated by this observation and value continuity, we define three rules for associating ATPs within the same call context. 
For an ATP $p_i=\langle\mathcal{E}_i,\mathcal{S}_i,\mathcal{D}_i\rangle \in \mathcal{P}_c$, let
$\mathbf{d}_{i,k} = \langle a^{d}_{i,k}, \tau^{d}_{i,k}, q^{d}_{i,k}, v^{d}_{i,k} \rangle \in \mathcal{D}_i$
denote its destination entry indexed by $k$, and let
$\mathbf{s}_{i,\ell} = \langle a^{s}_{i,\ell}, \tau^{s}_{i,\ell}, q^{s}_{i,\ell}, v^{s}_{i,\ell} \rangle \in \mathcal{S}_i$
denote its source entry indexed by $\ell$. 
Table~\ref{tab:atp-association-rules} summarizes the matching conditions for the three rules. 
When two or more ATPs are associated, the algorithm merges them into a single ATP. 
It combines their transfer sets and removes the matched source and destination entries because these entries become internal to the merged flow.
All unmatched entries are retained in the corresponding source or destination set.
The merged ATP then replaces the associated ATPs in
$\mathcal{P}_c$.

\begin{table}[!t]
\caption{Matching conditions for ATP association rules.}
\label{tab:atp-association-rules}
\centering
\footnotesize
\setlength{\tabcolsep}{3pt}
\renewcommand{\arraystretch}{1.2}

\begin{tabularx}{\columnwidth}{@{}
>{\raggedright\arraybackslash}p{0.13\columnwidth}
>{\raggedright\arraybackslash}p{0.21\columnwidth}
>{\raggedright\arraybackslash}X
@{}}
\toprule

\textbf{Rule}
&
\textbf{Candidate ATPs}
&
\textbf{Matching conditions}
\\

\midrule

\textbf{R1}
&
$p_i$, $p_j$
&
\(
p_i\prec p_j
\land
a^{d}_{i,k}=a^{s}_{j,\ell}
\land
\tau^{d}_{i,k}=\tau^{s}_{j,\ell}
\)
\\
\midrule

\textbf{R2}
&
$p_i$, $p_j$
&
\(
\begin{aligned}
&\operatorname{Adj}(p_i,p_j)
\land a^{d}_{i,k}=a^{s}_{j,\ell}\neq\varnothing\\
&\land \tau^{d}_{i,k}\neq\tau^{s}_{j,\ell}
\land v^{d}_{i,k}=v^{s}_{j,\ell}
\end{aligned}
\)
\\
\midrule

\textbf{R3-Mint}
&
$\langle p_{i_1},\ldots,p_{i_h}\rangle$,
$p_M$
&
\(
\begin{aligned}[t]
&a^s_M=\varnothing
\land
a^d_{i_1}=\cdots=a^d_{i_h}\neq\varnothing
\\
&\land
(\forall u\in\{1,\ldots,h\}:
\tau^d_{i_u}\neq\tau^s_M)
\\
&\land
(\forall u,w\in\{1,\ldots,h\},
u\neq w:
\tau^d_{i_u}\neq\tau^d_{i_w})
\\
&\land
\sum_{u=1}^{h}v^d_{i_u}=v^s_M
\end{aligned}
\)
\\
\midrule

\textbf{R3-Burn}
&
$p_B$,
$\langle p_{j_1},\ldots,p_{j_h}\rangle$
&
\(
\begin{aligned}[t]
&a^d_B=\varnothing
\land
a^s_{j_1}=\cdots=a^s_{j_h}\neq\varnothing
\\
&\land
(\forall u\in\{1,\ldots,h\}:
\tau^s_{j_u}\neq\tau^d_B)
\\
&\land
(\forall u,w\in\{1,\ldots,h\},
u\neq w:
\tau^s_{j_u}\neq\tau^s_{j_w})
\\
&\land
\sum_{u=1}^{h}v^s_{j_u}=v^d_B
\end{aligned}
\)
\\

\bottomrule
\end{tabularx}

\vspace{2pt}
\parbox{\columnwidth}{%
\footnotesize
\raggedright
\textit{Note:}
The relation $p_i\prec p_j$ indicates that $p_i$ precedes $p_j$ in $\mathcal{P}_c$, whereas
$\operatorname{Adj}(p_i,p_j)$ indicates that the two ATPs are adjacent regardless of their order.
For R3, the normal ATPs form a contiguous group immediately preceding the mint ATP or following the burn ATP.
}
\end{table}

\textbf{R1: Same-asset continuity.}
R1 captures the continuity of the same asset through a common
intermediate address. 
For each source entry $\mathbf{s}_{j,\ell}$, a preceding destination entry is considered compatible if the two entries satisfy the R1 matching conditions.
\textsc{AssociatePaths} first selects the closest preceding compatible destination entry with the same amount and merges the corresponding ATPs.
If no equal-amount entry exists, the algorithm selects the closest compatible entry satisfying $q^d_{i,k}>q^s_{j,\ell}$.
It splits the selected entry into a matched portion of amount $q^s_{j,\ell}$ and a residual portion of amount $q^d_{i,k}-q^s_{j,\ell}$. 
The matched portion is merged, while the residual portion remains in the destination set of the resulting ATP.
The USD value of the selected entry is divided between the two portions in proportion to their token amounts.
If no individual entry is sufficient, the algorithm accumulates compatible preceding entries from closest to farthest until their total amount reaches or exceeds $q^{s}_{j,\ell}$.
Only the ATPs contributing to this cumulative match are merged with $p_j$, and their aggregate is processed in the same manner.
If the cumulative amount never reaches $q^{s}_{j,\ell}$, no association is made.

\textbf{R2: Cross-asset continuity.}
R2 captures the continuity of value across asset conversions, such as DeFi swaps. 
Although the token type changes, the tracked value continues in another asset form.
When two adjacent ATPs satisfy the R2 matching conditions, the algorithm merges them, treating $\mathbf{d}_{i,k}$ and $\mathbf{s}_{j,\ell}$ as the matched entries.
Here, $p_i$ and $p_j$ denote the ATPs providing the destination and source entries, respectively. Their subscripts do not imply execution order, because the input and output transfers of an asset conversion may appear in either order in the execution trace.

\textbf{R3: Mint/burn boundary continuity.}
R3 treats token minting and burning as potential value transformation boundaries and captures value continuity across them.
For R3, we consider only candidate ATPs containing exactly one source entry and one destination entry.
We use \(\varnothing\) to denote the zero address.
An ATP is identified as a mint ATP $p_M$ if its source address is $\varnothing$ and as a burn ATP $p_B$ if its destination address is $\varnothing$.
A normal candidate ATP has neither a zero address source nor a zero address destination.

\emph{R3-Mint.}
Token minting creates new tokens and is commonly associated with deposits or asset locking in DeFi protocols.
For a mint ATP $p_M$, the algorithm examines a contiguous group of normal ATPs $\langle p_{i_1},\ldots,p_{i_h}\rangle$ immediately preceding $p_M$.
If the group satisfies the R3-Mint matching conditions, the algorithm merges these ATPs with $p_M$, treating the destination entries of the normal ATPs and the source entry of $p_M$ as the matched entries.

\emph{R3-Burn.}
Token burning removes tokens from circulation and may be accompanied by asset outflows during redemptions or liquidity removal in DeFi protocols.
For a burn ATP $p_B$, the algorithm examines a contiguous group of normal ATPs $\langle p_{j_1},\ldots,p_{j_h}\rangle$ immediately following $p_B$. 
If the group satisfies the R3-Burn matching conditions, the algorithm merges these ATPs with $p_B$, treating the destination entry of $p_B$ and the source entries of the normal ATPs as the matched entries.

\textsc{AssociatePaths} applies R1, R2, and R3 in this order.
After each association, it updates the ATP sequence and restarts the scan. The procedure terminates when a complete pass produces no new association.

\paragraph{Illustrative Example}
Fig.~\ref{fig:catt}(c) illustrates the recovery of asset-flow paths from a transaction $tx$ with five raw asset transfers, as shown in Fig.~\ref{fig:catt}(a).
Initially, each transfer $t_i$ is initialized as an atomic ATP $p_i$.
During a postorder traversal of the CATT, \textbf{R2} is first applied to associate $p_2$ and $p_1$ into $p'$, as their transfers share an intermediary address ($Addr_2$) and involve different tokens with the same USD value $v$.
Next, \textbf{R3-Burn} is applied to associate $p_3$ and $p_4$ into $p''$ based on the burn-related transfer at $Addr_3$.
Finally, \textbf{R1} is applied twice to combine $p'$ and $p''$ into $p'''$ and then associate $p'''$ with $p_5$, based on consecutive transfers of tokens B and C, respectively, through $Addr_3$.

\subsubsection{ATP-guided Laundering Network Expansion}
\label{sec4.5}
We use recovered ATPs to guide laundering network expansion, as detailed in Algorithm~\ref{alg:ml_fund_tracking}.
The algorithm initializes a queue with the ML source set \(S\) (Lines~\ref{alg:ml:init-graph}--\ref{alg:ml:init-queue}), where each pair \((a,b_a)\) records an address \(a\) and the block number \(b_a\) from which tracing begins. 
For each dequeued pair, it examines transactions containing asset outflows from $a$ after $b_a$ (Lines~\ref{alg:ml:dequeue}--\ref{alg:ml:query}), using $Visited$ to prevent repeated transaction processing. 
For each unvisited transaction \(tx\), it recovers its ATPs and selects those whose source sets contain \(a\), yielding \(\mathcal{P}_{tx}^{a}\) (Lines~\ref{alg:ml:unvisited}--\ref{alg:ml:paths}).  
For each selected ATP, the algorithm considers all destination addresses as downstream candidates and adds corresponding flow edges to $E_f$ (Lines~\ref{alg:ml:dest-loop}--\ref{alg:ml:add-edge}).
Destinations labeled as service providers in \(SP\), or those satisfying the stopping criterion, are added to \(V_t\) without further expansion (Lines~\ref{alg:ml:stop-test}--\ref{alg:ml:terminal}). 
The stopping criterion limits excessive expansion through highly active addresses and is instantiated according to the evaluation setting.
Other destinations are added to \(V_m\) and queued with the block number of \(tx\) (Lines~\ref{alg:ml:add-vm}--\ref{alg:ml:enqueue}). 
When the queue becomes empty, the algorithm returns the laundering graph \(G\), which records downstream ML flows and the tracing boundaries reached during expansion (Line~\ref{alg:ml:return}).

\begin{algorithm}[!t]
\caption{Path-Aware Laundering Network Expansion}
\label{alg:ml_fund_tracking}
\footnotesize
\setlength{\algorithmicindent}{0.75em}
\begin{algorithmic}[1]
\REQUIRE ML source set $S=\{(a,b_a)\}$,
         labeled service provider set $SP$
\ENSURE A recovered laundering graph $G=(V_m,V_t,E_f)$

\STATE Initialize an empty graph $G$,
       $V_m\gets\{a\mid(a,b_a)\in S\}$
       \label{alg:ml:init-graph}
       
\STATE Initialize queue $Q\gets S$,
       visited transaction set $Visited\gets\emptyset$
       \label{alg:ml:init-queue}

\WHILE{$Q$ is not empty}
    \STATE $(a,b_a)\gets Q.\textsc{PopFront}()$
    \label{alg:ml:dequeue}
    
    \STATE $\mathit{T}_a \gets
            \textsc{QueryOutflowTxsAfter}(a,b_a)$
            \label{alg:ml:query}

    \FOR{each $tx\in\mathit{T}_a$}
        \IF{$tx\notin Visited$}
        \label{alg:ml:unvisited}
        
            \STATE $Visited\gets Visited\cup\{tx\}$

            \STATE $\mathcal{P}_{tx}^{a}\gets
            \textsc{SelectOutgoingPaths}
            \bigl(\textsc{RecoverPaths}(tx),a\bigr)$
            \label{alg:ml:paths}

            \FOR{each $p\in\mathcal{P}_{tx}^{a}$}
                \FOR{each $\mathbf{d}\in\mathcal{D}_p$}
                \label{alg:ml:dest-loop}
                
                    \STATE $a_{\mathrm{dst}}\gets
                    \textsc{Address}(\mathbf{d})$

                    \STATE $E_f\gets E_f\cup
                    \{\langle a,a_{\mathrm{dst}},tx,p\rangle\}$
                    \label{alg:ml:add-edge}

                    \IF{$\textsc{IsKnownService}
                    (a_{\mathrm{dst}},SP)$
                    \OR $\textsc{ShouldStop}(a_{\mathrm{dst}})$}
                    \label{alg:ml:stop-test}

                        \STATE $V_t\gets V_t\cup
                        \{a_{\mathrm{dst}}\}$
                        \label{alg:ml:terminal}

                    \ELSE
                        \STATE $V_m\gets V_m\cup
                        \{a_{\mathrm{dst}}\}$
                        \label{alg:ml:add-vm}

                        \STATE $Q.\textsc{PushBack}
                        \bigl(
                        (a_{\mathrm{dst}},
                        \textsc{BlockNumber}(tx))
                        \bigr)$
                        \label{alg:ml:enqueue}
                    \ENDIF
                \ENDFOR
            \ENDFOR
        \ENDIF
    \ENDFOR
\ENDWHILE

\STATE \textbf{return} $G=(V_m,V_t,E_f)$
\label{alg:ml:return}
\end{algorithmic}
\end{algorithm}

\subsection{Cross-chain Deposit Transaction Identification}
\label{sec5}
Typically, users first initiate a deposit transaction on the source chain to enter the cross-chain workflow.
By identifying such transactions during laundering tracing, we can capture the cross-chain transfer intent of illicit actors and provide investigators with observable evidence to support subsequent destination-chain analysis.
Our goal is therefore to determine whether an SP-related transaction associated with $V_t$ carries the tracked illicit assets into a cross-chain bridge.
However, SP-related transactions collected during laundering tracing are highly heterogeneous, covering CEXs, single-chain DeFi protocols, mixers, and cross-chain bridges.
Identification based on manually maintained contract lists or function identifiers may fail to cover diverse and evolving bridge protocols, limiting its generality.

Based on the cross-chain workflow described in Section~\ref{bridge_background}, we observe that, despite diverse bridge implementations, source-chain deposit transactions generally can be characterized from two complementary views. 
The first, the asset-flow view, focuses on how user' assets are transferred and processed within a deposit transaction.
The second, the execution-evidence view, focuses on how bridge operations expose cross-chain processing information for off-chain components.
For the execution-evidence view, we further observe a protocol-level distinction between cross-chain bridge deposits and single-chain DeFi interactions.
Single-chain DeFi transactions often maintain user-related states in contract storage to support later withdrawals, claims, or settlement, whereas bridge deposits commonly expose user-related cross-chain information through event logs that can be monitored by off-chain components such as relayers or validators. Therefore, storage and event signals provide complementary signals for distinguishing cross-chain deposits from single-chain DeFi interactions.
Based on this observation, FPTracer extracts four groups of features from the asset-flow view and the execution-evidence view, as shown in Table \ref{tab:feature_groups}, and applies a learning-based classifier to determine whether an SP-related transaction carries the tracked illicit assets into a cross-chain bridge.

\begin{table}[!t]
\caption{Four groups of features.}
\label{tab:feature_groups}
\centering
\footnotesize
\setlength{\tabcolsep}{3pt}
\renewcommand{\arraystretch}{1.15}

\begin{tabular}{
>{\raggedright\arraybackslash}p{0.15\columnwidth}
>{\raggedright\arraybackslash}p{0.1\columnwidth}
>{\raggedright\arraybackslash}p{0.22\columnwidth}
>{\raggedright\arraybackslash}p{0.45\columnwidth}
}

\toprule
\textbf{Feature}
& \textbf{Level}
& \textbf{Source}
& \textbf{Description}
\\
\midrule

Path distribution
& Trans\-action
& ATP set
& Path count, single-path flag and normalized HHI over ATP value shares.
\\

Asset flow
& Path
& ATP set
& Path length, circular-path flag, destination role, destination relation, last-hop asset type, asset conversion count.
\\

Destination behavior
& Path
& ATP set and historical address records
& Historical destination behavior, including transaction counts (total/incoming/outgoing), transfer volume (total/inflow/outflow), and temporal behavior.
\\

Execution state
& Path
& CATT, ATP set, event logs, storage changes
& Features around the path anchor, including storage update count, user-related storage update count, and event count.
\\
\bottomrule

\end{tabular}
\end{table}

\textbf{Path Distribution Features} describe the transaction-level distribution of the recovered ATPs. They include the path count and path-value distribution. 
We measure value distribution using the normalized Herfindahl--Hirschman Index (HHI). 
For a candidate transaction, let \(n = |\mathcal{P}|\) denote the number of ATPs. 
For each path \(p \in \mathcal{P}\), let \(p.v_{usd}\) denote its USD value and define its value share as \(s_p = p.v_{usd} / \sum_{q \in \mathcal{P}} q.v_{usd}\).
The normalized HHI is defined as:
\begin{equation}
\mathrm{HHI}^{norm} =
\frac{\sum_{p \in \mathcal{P}} s_p^2-\frac{1}{n}}
{1-\frac{1}{n}},
\quad n>1.
\end{equation}
For \(n=1\), we set \(\mathrm{HHI}^{norm}=1\).
A larger \(\mathrm{HHI}^{norm}\) indicates that the transferred value is concentrated in one or fewer dominant paths.

\textbf{Asset Flow Features} describe how the traced value moves within an ATP and where the path ends. They capture three aspects: path structure, such as path length and a flag indicating whether a path ends at its start address; path destination semantics, including whether the destination address is a CA, EOA, or $\varnothing$, and its relation to \texttt{tx.from} and \texttt{tx.to}; and asset semantics, such as the last-hop asset type and the number of asset conversions. 

\textbf{Destination Behavior Features} characterize the historical behavior of the path destination before the target transaction. Following prior address-behavior profiling methods~\cite{tharani2024unified,wu2026ettracker}, we summarize the destination's transaction activity and transfer volume using total, incoming, and outgoing statistics. We further capture recent temporal behavior within a 30-day window.

\begin{lstlisting}[
    style=soliditycompact,
    caption={Example of user-related storage updates through Solidity mappings.},
    label={lst:user_related_storage}
]
...... // omitted state variables
// mapping entry
mapping(address => uint256) private userDebts; 
...... // omitted state variables
function _enqueueDebt(address account, uint256 amount) internal {
    ......
    // user-related storage update
    userDebts[account] += amount; 
    ......
}
\end{lstlisting}

\textbf{Execution State Features} capture event and storage signals in the execution context associated with an ATP.
For each ATP, we select the call node associated with its last-hop transfer in the CATT as the analysis anchor. We then extract event-level and storage-level features from the execution context of this anchor.
The event-level feature counts event logs emitted within the execution subtree rooted at the anchor, excluding standardized token events such as \texttt{Transfer} and \texttt{Approval}.
Storage-level features capture storage changes within the same execution context, with particular attention to user-related updates.
User-related states are commonly implemented as mapping entries indexed by user addresses. 
As shown in Listing~\ref{lst:user_related_storage}, \texttt{userDebts} is a \texttt{mapping(address => uint256)} variable that uses a user account as the key. It maintains debt states for individual users. 
The statement \texttt{userDebts[account] += amount} modifies the mapping entry associated with \texttt{account} and may cause a user-related storage update. 
If the base slot of \texttt{userDebts} is \(b\), the storage location of \texttt{userDebts[account]} is computed as: $\mathrm{keccak256}(h(account) \,\|\, b)$, where \(\|\) denotes concatenation and \(h(account)\) denotes the 32-byte encoding of the address key~\cite{solidity_storage_layout}. 
Based on this rule, we identify potential user-related storage updates using heuristic matching.
For each anchor, we collect the changed storage slots and enumerate a bounded range of possible base slots.
For each candidate base slot, we derive the storage location using the relevant user address as the mapping key.
If a candidate storage location appears in the changed storage slots, we mark it as a user-related storage update.  
This heuristic does not recover the complete storage schema, but provides a lightweight signal of persistent user-associated state changes.

\textbf{Feature Integration.}
The four feature groups operate at different granularities. Path distribution features are computed at the transaction level, whereas asset flow, destination behavior, and execution state features are extracted for individual ATPs. We aggregate the path-level features using each ATP's value share as the weight and concatenate the resulting representations with the transaction-level features to obtain a fixed-length feature vector for classification.

\subsection{Mixer Interaction Analysis}
\label{sec6}
Mixers interrupt deterministic asset tracing on Ethereum, as the withdrawal corresponding to the tracked illicit assets cannot be directly determined.
However, the laundering network $G$ recovered in the preceding stage provides additional evidence for downstream analysis. 
In particular, the address set \(V_m\), which contains addresses strongly associated with the illicit actors, can help identify potential withdrawal-side laundering transactions.
In this section, we first identify Tornado Cash deposit transactions from SP-related transactions and then leverage the preceding laundering evidence to derive candidate withdrawal-side transactions, providing additional leads for downstream forensic analysis.
These candidates are supported by observable laundering evidence but do not establish deterministic correspondences between individual deposits and withdrawals.

\textbf{Identification of Mixer Deposits.}
A deposit transaction can be executed either directly through a pool contract or indirectly through an auxiliary contract, such as a router or proxy. In both cases, the deposited asset ultimately reaches a Tornado Cash pool. We therefore identify deposits among SP-related transactions based on ATP destinations.
We first collect known Tornado Cash pool contracts from the official documentation~\cite{tornadocash_smart_contracts}.
Given the set of SP-related transactions $\mathcal{T}_{SP}$, we mark $tx \in \mathcal{T}_{SP}$ as a deposit if at least one of its ATPs terminates at a pool contract and the transferred asset type and amount match the pool's fixed denomination. The resulting deposit set is denoted by $\mathcal{T}_{d}$.

\textbf{Withdrawal-side Laundering Analysis.}
After identifying deposits, we conduct withdrawal-side analysis using evidence from the preceding laundering network.
For each pool $c$, we collect the deposits in $\mathcal{T}_{d}$ associated with $c$ and define an analysis window $[t_{\min}^{c}, t_{\max}^{c}+\Delta_w]$, where $t_{\min}^{c}$ and $t_{\max}^{c}$ denote the earliest and latest deposit times for pool $c$, respectively, and $\Delta_w$ extends the window to account for delayed withdrawals.
We collect all withdrawals from the pool $c$ within this window as the withdrawal analysis set.
Based on the mechanism of Tornado Cash described in Section~\ref{tornado_background}, we design the following rules to identify candidate withdrawal-side transactions.

\emph{Rule 1:} If the \emph{withdrawer} appears in $V_m$, or has prior asset-transfer interactions with addresses in $V_m$, we retain the transaction as a withdrawal-side candidate.

\emph{Rule 2:} After a withdrawal, we follow the downstream asset flows from the \emph{withdrawer} for up to two transfer hops. If any recipient within these two hops appears in $V_m$, we retain the withdrawal transaction as a candidate.

\emph{Rule 3:} If \(\texttt{tx.from}\) is not a \emph{relayer} and either appears in \(V_m\) or has previously received assets from addresses in \(V_m\), we retain the transaction as a candidate.

\emph{Rule 4:} If a \emph{depositor} deposits into multiple pools and a \emph{withdrawer} later withdraws from the same set of pools with matching numbers of deposits and withdrawals in each pool, we treat this cross-pool behavioral consistency as supporting evidence and retain the corresponding withdrawals as candidates.

\section{Evaluation}
\label{sec7}

We evaluate FPTracer by answering the following research questions (RQs):

\begin{itemize}
    \item \textbf{RQ1:} How accurately does FPTracer recover underlying asset-flow paths in laundering transactions?

    \item \textbf{RQ2:} How effective is FPTracer in laundering network expansion?

    \item \textbf{RQ3:} How accurately does FPTracer identify cross-chain deposit transactions during laundering tracing?

    \item \textbf{RQ4:} Can FPTracer effectively identify post-mixer laundering flows?

\end{itemize}

\subsection{Evaluation Setup}

\textbf{Baseline.}
We compare FPTracer with three representative open-source tracing approaches: XBlockFlow~\cite{Wu2024}, and AMLGuard~\cite{wu2026tracingshadowsautomatictracking}. 
XBlockFlow propagates illicit fund flows using taint-based tracing.
AMLGuard performs semantic-aware laundering tracing based on inferred transaction semantics and further supports cross-chain intent identification. 
We compare both two approaches with FPTracer for laundering-network expansion in RQ2, while AMLGuard, the only approach supporting cross-chain intent identification, is further compared with FPTracer in RQ3.
AMLGuard is open source, but its released artifact does not include the DeFi operation knowledge base required by its RAG-based inference component. We therefore evaluate the executable rule-based workflow provided in the released artifact, denoted as AMLGuard-Public.
To the best of our knowledge, no existing methods are directly comparable under the task settings of RQ1 and RQ4. We therefore evaluate RQ1 against manually annotated ground truth and validate RQ4 using external public evidence.

\textbf{Dataset.} 
To assess the practical effectiveness of FPTracer, We collect 20 real-world laundering incidents, that occurred between 2019 and 2025 from public reports released by reputable blockchain security firms, and use them as our case-level evaluation set, denoted by \(D_{case}\).
For each case, we obtain Ethereum execution records, including calls, transfers, event logs, and storage changes, through Bitquery~\cite{bitquery_docs}. These records can also be obtained from archive nodes or execution replay.
To support quantitative evaluation, we further use the annotated AML dataset released by AMLGuard, containing labeled laundering traces for 82 real-world incidents, to construct the evaluation benchmarks used in subsequent experiments. 
For RQ1, we sample laundering transactions from this dataset and manually annotate their intra-transaction asset-flow paths as ground truth to form \(B_{path}\). For RQ2, we select the ten cases in \(D_{case}\) that overlap with the AMLGuard dataset and use their annotated laundering traces as reference annotations to construct \(B_{trace}\).
Separately, for RQ3, we construct a transaction-level benchmark \(B_{bridge}\) to train and evaluate the cross-chain deposit classifier.
For the case-level bridge analysis in RQ3 and the mixer analysis in RQ4, no comprehensive public benchmarks are available. We therefore use publicly reported evidence and manual verification as reference evidence, with the detailed construction and validation procedures described in the corresponding RQs.

\subsection{RQ1: Results of Intra-transaction Path Recovery}
\label{sec7.3}
The recovery of asset-flow paths in transaction serves as the foundation for the subsequent expansion of the laundering network and service analysis. 
We therefore evaluate its recovery accuracy on laundering transactions.
Specifically, we randomly sample 300 laundering transactions containing at least two asset transfers from the AMLGuard dataset and manually annotate their asset-flow paths as ground truth, forming $B_{path}$.

A transaction is considered exactly recovered only if its recovered path set exactly matches the manual annotation, without missing or spurious paths.
FPTracer exactly recovers 298 of 300 transactions, yielding a Transaction Exact Recovery Rate (TER) of 99.33\%.
We manually analyze the remaining two transactions and find that both failures arise from uncommon protocol execution design.
For example, Sky LitePSM uses different contracts within the same protocol to handle the input and output assets. 
As a result, the destination of the input transfer differs from the source of the output transfer, violating the shared intermediate address condition required by \textbf{R2} and causing the recovered path to be fragmented.

\subsection{RQ2: Results of Laundering Network Tracing}
\label{RQ2}
In this section, we evaluate the effectiveness of FPTracer in laundering tracing on $B_{trace}$. 
We adopt Recall and Destination Recall (D-Recall) as evaluation metrics.
Recall measures the proportion of labeled laundering addresses recovered by a method, while D-Recall measures the proportion of labeled laundering destinations reached during tracing. 

As shown in Table~\ref{tab:laundering_tracing}, FPTracer achieves a Recall of 98.6\% and a D-Recall of 94.2\%, indicating effective downstream tracing based on recovered intra-transaction asset-flow paths.
XBlockFlow achieves lower tracing coverage, which is attributed to its reliance on transfer relationships without explicitly reconstructing continuous asset flow paths within complex Ethereum transactions.
AMLGuard-Public achieves higher coverage by incorporating DeFi operation semantics, but its rule-based semantic recognition fail for transactions involving mint/burn transformations or requiring execution-context constraints.

\begin{table}[!t]
\centering
\footnotesize
\caption{Comparison of tracing effectiveness.}
\label{tab:laundering_tracing}
\begin{tabular}{lcc}
\toprule
\textbf{Method} & \textbf{Recall} &
\textbf{D-Recall}   \\
\midrule
XBlockFlow~\cite{Wu2024}  & 90.2\% & 84.7\%  \\
AMLGuard-Public~\cite{wu2026tracingshadowsautomatictracking}
& 95.3\% & 87.0\%  \\
\midrule
\textbf{FPTracer}      & \textbf{98.6\%} & \textbf{94.2\%}  \\
\bottomrule
\end{tabular}
\end{table}

\subsection{RQ3: Effectiveness of Cross-Chain Deposit Transaction Identification}
We evaluate cross-chain deposit identification at two complementary levels.
First, we evaluate the transaction-level identification performance of \textbf{M2} on the labeled benchmark $B_{bridge}$.
Second, we examine FPTracer's effectiveness in identifying bridge usage during laundering tracing on $D_{case}$.

\subsubsection{Evaluation of M2.}
We construct a labeled benchmark $B_{bridge}$ containing 8,533 cross-chain deposit transactions from 28 bridges protocols and 8,441 non-deposit transactions.
We collect deposit transactions from public datasets released by prior studies~\cite{yan2025empirical,connector2025} and from selected bridges listed by Chainspot~\cite{chainspot_bridges} and DeFiLlama~\cite{defillama_bridges}. For the latter, we obtain bridge contract addresses from official documentation and extract deposit transactions using protocol-specific function selectors and event signatures. The resulting transactions are then deduplicated.
We additionally collect non-deposit transactions, including CEX-related transactions, bridge-related non-deposit interactions, and single-chain DeFi interactions, to better reflect confounding transactions encountered during laundering tracing.

\begin{table*}[!t]
\caption{Classification performance under two evaluation settings.}
\label{tab:cross_chain_identification}
\centering
\footnotesize
\setlength{\tabcolsep}{4pt}
\renewcommand{\arraystretch}{1.15}

\begin{tabular*}{\textwidth}{@{\extracolsep{\fill}}lcccccccc@{}}
\toprule
\multirow{2}{*}{\textbf{Classifier}}
& \multicolumn{4}{c}{\textbf{Address-Disjoint}}
& \multicolumn{4}{c}{\textbf{Bridge-Disjoint}} \\
\cmidrule(lr){2-5}
\cmidrule(lr){6-9}
& \textbf{Precision}
& \textbf{Recall}
& \textbf{Accuracy}
& \textbf{F1-score}
& \textbf{Precision}
& \textbf{Recall}
& \textbf{Accuracy}
& \textbf{F1-score} \\
\midrule

DT
& 98.77\%$\pm$0.82
& 89.26\%$\pm$8.64
& 94.06\%$\pm$4.34
& 93.59\%$\pm$4.93
& 99.04\%$\pm$1.60
& 92.26\%$\pm$8.56
& 95.67\%$\pm$4.15
& 95.30\%$\pm$4.67 \\

RF
& \textbf{99.10\%$\pm$0.57}
& 98.14\%$\pm$1.11
& 98.61\%$\pm$0.74
& 98.61\%$\pm$0.73
& 99.20\%$\pm$1.52
& 98.08\%$\pm$8.17
& 98.63\%$\pm$4.10
& 98.42\%$\pm$5.02 \\

SVC
& 98.52\%$\pm$1.46
& 97.80\%$\pm$2.51
& 98.14\%$\pm$1.38
& 98.13\%$\pm$1.40
& 99.23\%$\pm$1.55
& 97.67\%$\pm$3.66
& 98.45\%$\pm$1.96
& 98.40\%$\pm$2.05 \\

XGBoost
& 98.41\%$\pm$1.31
& \textbf{98.92\%$\pm$1.19}
& \textbf{98.65\%$\pm$1.06}
& \textbf{98.66\%$\pm$1.04}
& 98.83\%$\pm$1.98
& 97.23\%$\pm$8.53
& 98.05\%$\pm$4.57
& 97.82\%$\pm$5.40 \\

\bottomrule
\end{tabular*}
\end{table*}

To validate the overall effectiveness of method, we employ four standard classifiers: Decision Tree (DT), Random Forest (RF), Support Vector Classifier (SVC) and XGBoost. Since several features depend on path destinations, we enforce address isolation to avoid feature leakage. Specifically, transactions sharing the same destination address are assigned exclusively to either the training or test set. The resulting train/test ratio is approximately 8:2. 
We repeat the split using different random seeds and report the mean and standard deviation of the results. 
As shown in Table~\ref{tab:cross_chain_identification}, all classifiers except DT achieve F1 scores above 98\% , indicating that the feature representation provides useful discriminative signals and is not strongly dependent on a specific classifier.
XGBoost achieves the highest recall, accuracy, and F1 score while maintaining high precision. We therefore adopt XGBoost as the classifier for the subsequent case-level analysis.

To further evaluate generalization to unseen bridge protocols, we conduct repeated bridge-disjoint experiments. In each run, we randomly select 3 bridge protocols and reserve all their deposit transactions for testing, while deposits from the remaining bridges are used for training. 
We repeat this process and ensure that every bridge is included in the test set at least once.
All classifiers except DT retain high
performance, with F1 scores above 97\%, as reported in Table~\ref{tab:cross_chain_identification}. 
However, the standard deviations are larger than those under the address-disjoint setting. This variation indicates that generalization performance is more sensitive to the choice of held-out bridges, reflecting differences in bridge designs.

\subsubsection{Evaluation of FPTracer's Cross-Chain Deposit Identification.}
We further evaluate how effectively FPTracer identifies cross-chain deposit transactions during laundering tracing on $D_{case}$.
We select \emph{AMLGuard-Public} as the comparison method because it explicitly supports the identification and tracing of cross-chain laundering activities.
Because complete ground truth for bridge usage is lack for the 20 laundering cases, we manually reviewed public incident reports from blockchain security firms. Explicit bridge usage is reported for five cases.
These reports are not exhaustive, but provide external evidence for evaluating the recovery of known bridge usage.
We use a case--bridge pair as the unit of comparison. A case--bridge pair denotes a unique combination of a laundering case and a bridge protocol. Multiple transactions involving the same bridge within one case are counted as one pair.
Reported-pair Coverage (RPC) is used to measure the proportion of reported case--bridge pairs recovered by each method.

\begin{table}[!t]
\caption{Comparison of AMLGuard and FPTracer in identifying cross-chain bridge usage.}
\label{tab:bridge_quantitative_comparison}
\centering
\footnotesize
\setlength{\tabcolsep}{2pt}
\renewcommand{\arraystretch}{1.15}

\begin{tabularx}{\columnwidth}{@{}
>{\raggedright\arraybackslash}p{1.9cm}
*{3}{>{\centering\arraybackslash}p{0.15\columnwidth}}
>{\centering\arraybackslash}X
@{}}

\toprule
\textbf{Method}
&
\makecell[c]{\textbf{Identified}\\\textbf{Pairs}}
&
\makecell[c]{\textbf{Reported}\\\textbf{Pairs}}
&
\textbf{RPC}
&
\makecell[c]{\textbf{Additional}\\\textbf{Verified Pairs}}
\\
\midrule

AMLGuard-Public
& $6$
& $3 $
& $27.27\%$
& $3$ \\

FPTracer
& $41$
& $11 $
& $100\%$
& $30$ \\

\bottomrule
\end{tabularx}

\vspace{2pt}
\parbox{\columnwidth}{%
    \footnotesize
    \raggedright
    \textit{Note:}
    \emph{Identified Pairs} denotes the total number
    of case-bridge pairs identified by each method.
    \emph{Reported Pairs} denotes the identified pairs
    that are documented in the collected public reports.
    \emph{Additional Verified Pairs} denotes pairs
    not mentioned in the collected reports but
    confirmed through manual validation.
}

\end{table}

FPTracer identifies bridge usage in 10 cases.
Table~\ref{tab:bridge_quantitative_comparison} summarizes the comparison with AMLGuard-Public. 
FPTracer recovers all reported pairs, achieving 100\% RPC, whereas AMLGuard-Public misses several reported pairs.
Our manual inspection indicates that the missed pairs mainly involve bridge contracts or function signatures not covered by AMLGuard-Public and transactions routed through proxy contracts.
FPTracer further identifies 30 additional pairs not mentioned in the collected public reports.
We manually validate them using multiple sources of evidence: 1)tracing the deposited assets back to labeled ML addresses; 2) cross-checking the transactions using bridge explorers, where available; 3) verifying the involved contracts and examining function signatures and protocol-specific events for evidence of cross-chain deposit semantics. 
All 30 additional pairs were supported by the collected on-chain evidence.

\subsection{RQ4: Results of Mixer Interaction Analysis}
\label{sec7.5}

FPTracer identifies Tornado Cash deposits in 18 of the 20 cases, comprising 4,253 deposit transactions with a total of 405,374.9 ETH.
The identified deposits involve all four ETH-denominated pools: 0.1 ETH, 1 ETH, 10 ETH, and 100 ETH. 
The 100-ETH pool dominates these interactions, accounting for 4,039 deposits (94.97\%) and 403,900 ETH (99.64\%) of the total deposited amount.
These results show that Tornado Cash constitutes a recurring tracing boundary in our laundering cases, highlighting the importance of explicitly identifying mixer interactions during laundering tracing.

Among the 18 cases involving Tornado Cash deposits, FPTracer identifies withdrawal-side candidates in seven cases using rules described in Section~\ref{sec6}.
Since Tornado Cash cryptographically breaks the observable link between deposits and withdrawals, no explicit on-chain ground truth reveals the true deposit--withdrawal correspondences. Existing studies therefore rely on auxiliary information or behavioral leakage to infer potential associations between mixer deposits and withdrawals~\cite{wang2023zkpmixers,Du2024}. 
To further corroborate the identified results, we extensively search publicly available security reports and incident analyses for all seven cases.
We find sufficiently specific withdrawal-side evidence for two cases: KucoinHacker and LiquidHacker.
External evidence corroborates 222 of the 442 candidates in KucoinHacker and 47 of the 82 candidates in LiquidHacker.
The external evidence provides positive corroboration for a subset of the candidates identified using preceding laundering evidence.
The remaining candidates are not treated as false positives because the available public evidence is not exhaustive.
Overall, although Tornado Cash prevents deterministic tracing across the mixer boundary, FPTracer can provide investigative evidence for potential withdrawal-side laundering activities and support subsequent forensic analysis.

\section{Related Work}
\label{sec2.5}
\textbf{Anti-Money Laundering on Ethereum.}
The pseudonymous and permissionless nature of cryptocurrencies create favorable conditions for money laundering, which has attracted increasing research attention~\cite{DecentralizedML,trozze2023degens,zhou2023visual,xiong2026understanding}.
Recent studies have proposed various methods to address on-chain AML challenges.
Several studies~\cite{farrugia2020detection,liu2023gtn2vec,stableAML,fu2026flowshield} employ machine learning to detect illicit accounts and transactions by modeling account behavior and transaction graph.
DenseFlow~\cite{lin2024denseflow} and GMPA~\cite{Li2025} further model money laundering as group behavior and use subgraph mining to detect laundering groups.
Another line of research focuses on tracing illicit assets after security incidents.
Wu et al.~\cite{Wu2024} trace downstream laundering activities from heists and analyze laundering behaviors.
ETTracker~\cite{wu2026ettracker} performs UMDG-based asset tracking and employs a GNN to identify unlabeled service providers.
AMLGuard~\cite{wu2026tracingshadowsautomatictracking} abstracts low-level token movements into transaction semantics and uses the inferred semantics to guide iterative laundering tracing.

\textbf{Identification of Cross-chain Transactions.} 
Identifying cross-chain transactions~\cite{zheng2026sok} provides a basis for analyzing bridge activity~\cite{hu2025piecing,yan2025empirical}.
Existing studies identify such transactions using official records or APIs~\cite{yousaf2019tracing,zongyang2021cltracer} or by matching known function and event signatures~\cite{hu2025piecing,augusto2025xchainwatcher,wu2026tracingshadowsautomatictracking}. 
These approaches therefore depend on available bridge interfaces or maintained protocol information.
Another line of research~\cite{connector2025,abctracer2025} extracts cross-chain semantics from bridge execution signals and employs learning models for transaction classification.
These methods extract asset-transfer semantics from transfer graphs of transactions and message-passing semantics from event logs or calldata.
These methods are generally evaluated on transactions associated with known bridge protocols or contracts.
However, in laundering tracing, bridge deposits must be distinguished from heterogeneous service-provider interactions along the tracked fund flows.

\textbf{Mixer Analysis.}
Existing studies have investigated cryptocurrency mixers from the perspectives of anonymity, transaction linkability, and illicit usage.
Wang et al.~\cite{wang2023zkpmixers} empirically show that user behaviors can reduce the effective anonymity sets of mixers, while Béres et al.~\cite{Bres2021} and Tang et al.~\cite{tang2022analysis} analyze Tornado Cash linkability using heuristic rules.
Recent learning methods~\cite{Du2024,wang2025mixlinker,wu2025unveiling} further model mixing interactions using graph representations to infer associations between deposits and withdrawals.
Liu et al.~\cite{liu2026evasion} empirically evaluate Tornado Cash usage under blockchain sanctions and examine its continued use in Ethereum security incidents.

\section{Conclusion}
\label{sec8}
In this paper, we presented FPTracer, a path-aware framework for tracing illicit asset flows through on-chain services on Ethereum. FPTracer recovers intra-transaction asset-flow paths based on the CATT and uses these paths to guide laundering network expansion. 
It further identifies transactions exhibiting cross-chain intent and analyzes mixer interactions to identify potential withdrawal-side laundering activities.
Evaluation on 20 real-world laundering cases demonstrates the effectiveness of FPTracer and highlights the value of combining path-aware tracing with cross-chain bridge and mixer analysis for more comprehensive laundering analysis on Ethereum.

\bibliographystyle{IEEEtran}
\bibliography{references}

\end{document}